\documentstyle[12pt]{article}

\def\Journal#1#2#3#4{{#1} {\bf #2}, #3 (#4)}

\def\NPB{{\em Nucl. Phys.} B}
\def\PLB{{\em Phys. Lett.}  B}
\def\PRL{\em Phys. Rev. Lett.}
\def\PRD{{\em Phys. Rev.} D}
\def\ZPC{{\em Z. Phys.} C}

\def\be{\begin{equation}}
\def\ee{\end{equation}}
\def\bea{\begin{eqnarray}}
\def\eea{\end{eqnarray}}
\begin{document}

\centerline{\large \bf PQCD FACTORIZATION OF TWO-BODY $B$ DECAYS}
\vskip 1.0cm

\centerline{Hsiang-nan Li}
\vskip 0.3cm

\centerline{Institute of Physics, Academia Sinica,
Taipei, Taiwan 115, ROC}
\vskip 0.3cm

\centerline{Department of Physics, National Cheng-Kung University,}

\centerline{Tainan, Taiwan 701, ROC}

\vskip 1.0cm

\centerline{\bf abstract}
\vskip 0.3cm

I review the known approaches to two-body nonleptonic
$B$ meson decays, including factorization assumption,
modified factorization assumption, QCD factorization, and
perturbative QCD factorization. Important phenomenological
aspects of these approaches are emphasized.

\section{Introduction}

Two-body nonleptonic $B$ meson decays are a challenging
subject for both theorists and experimentalists. These modes are
complicated because of nonperturbative QCD dynamics, and important,
because measurements of their CP violation reveal the information of
the unitarity angles. In this talk I will review the known theoretical
approaches to two-body nonleptonic $B$ meson decays, which 
include factorization assumption (FA), modified factorization
assumption (MFA), QCD factorization (QCDF), and perturbative QCD (PQCD)
factorization.

\section{Factorization Assumption}

The conventional approach to two-body nonleptonic $B$ meson decays is
based on FA \cite{BSW}, in which nonfactorizable and annihilation
contributions are neglected and final-state-interaction effects are
assumed to be absent \cite{CHL}. Factorizable contributions are expressed
as products of Wilson coefficients, meson decay constants, and hadronic
form factors, which are then parametrized by models. For example, the
amplitude of the decay $B\to\pi\pi$ can be written as
\begin{eqnarray}
A(B\to\pi\pi)&=&C(\mu)\langle \pi\pi|O(\mu)|B\rangle\;,
\nonumber\\
& \approx &C(\mu)f_\pi\langle \pi|(\bar qb)_{V-A}|B\rangle\;,
\label{fa}
\end{eqnarray}
where $C$ is the Wilson coefficient, $O$ the four-fermion operator,
$f_\pi$ the decay constant, $(\bar q b)_{V-A}$ the $V-A$ weak current,
$\mu$ the renormalization scale, and the matrix element in the second
line the $B\to\pi$ transition form factor $F^{B\pi}$. Under the above
approximation, FA is simple and provides qualitative estimation of
branching ratios of various two-body nonleptonic $B$ meson decays.

However, there exist several serious theoretical drawbacks in FA. First,
FA breaks the scale independence of decay amplitudes, which are physical
quantities. Meson decay constants and form factors, being measurable, are
scale-independent, while Wilson coefficients are scale-dependent as
indicated in Eq.~(\ref{fa}). Hence, decay amplitudes, expressed as their
products, become scale-dependent.

Second, nonfactorizable amplitudes are not always negligible. It has been
known that the decay modes, whose factorizable contributions arise from
the internal $W$-emission with the small Wilson coefficient
$a_2=C_1+C_2/N_c$, are dominated by nonfactorizable contributions. The
ratio of the $B^0_d\to D^+\pi^-$ and $B^+\to D^0\pi^+$ branching ratios
and the $B\to J/\psi K^{(*)}$ decays are the examples \cite{CL,YL}.

Third, the evaluation of strong phases is ambigious in FA. Strong phases
are crucial, since they are related to CP asymmetries $A_{CP}$ in
two-body nonleptonic $B$ meson decays:
\begin{eqnarray}
A_{CP}\propto \sin\delta\sin\phi\;,
\end{eqnarray}
where $\delta$ represents the strong phase. To extract the unitarity
angle $\phi$ from the data of $A_{CP}$, $\delta$ must be determined
unambigiously. In FA strong phases arise from the Bander-Silverman-Soni
(BSS) mechanism (the charm loop) \cite{BSS}, which are proportional to
\begin{eqnarray}
\int du u(1-u)\theta (u(1-u)q^2-m_c^2)\;,
\label{bss}
\end{eqnarray}
$q$ being the external momentum flowing through the charm loop and $m_c$
the charm quark mass. Since
$q^2$ is unknown, the above integral has large uncertainty.

\section{Modified Factorization Assumption}

To improve the theoretical approach to two-body nonleptonic $B$ meson
decays, FA has been modified. It has been proposed to extract the $\mu$
dependence of the matrix element $\langle \pi\pi|O(\mu)|B\rangle$
before applying FA \cite{Ali,CT98}. The procedure is performed as follows:
\begin{eqnarray}
A(B\to\pi\pi)&=&C(\mu)g(\mu)\langle \pi\pi|O|B\rangle_{\rm tree}\;,
\nonumber\\
& \approx &C_{\rm eff} f_\pi\langle \pi|(\bar qb)_{V-A}|B\rangle\;,
\label{mfa}
\end{eqnarray}
where $g(\mu)$ represents the $\mu$ dependence of the matrix element and
the effective Wilson coefficient $C_{\rm eff}\equiv C(\mu)g(\mu)$ is
scale-independent. The extraction of $g(\mu)$ invloves the one-loop
corrections to the four-quark vertex, which are, however, infrared
divergent. If the infrared divergences are regulated by considering
external quarks off-shell by $-p^2$, the decay amplitude becomes
gauge-dependent. That is, $C_{\rm eff}$ in fact depends on the infrared
cutoff $-p^2$ and a gauge parameter $\lambda$. Therefore, the problem
of the scale dependence in FA is not really solved in MFA, but just
replaced by the one of gauge dependence.

In MFA nonfactorizable contributions are included via the parameter
so-called effective color number $N_c^{\rm eff}$. For example, $a_2$
associated with the internal $W$-emission is written as
\begin{eqnarray}
a_2=C_1+\frac{C_2}{N_c^{\rm eff}}\;.
\end{eqnarray}
By varying $N_c^{\rm eff}$, one can obtain better fit to the data of
two-body nonleptonic $B$ meson decays. However, $N_c^{\rm eff}$ is
obviously process-dependent, such that the predictive power of MFA
is weak. The above prescription also implies that nonfactorizable
contributions are real, an assumption which is certainly not general
enough in the viewpoint of parametrization. In MFA strong phases still come
from the BSS mechanism, which are ambigious as explained in the previous
section.

\section{QCD Factorization}

Recently, Beneke {\it et al.} proposed the QCDF approach to two-body
nonleptonic $B$ meson decays \cite{BBNS}, in which the above drawbacks
of FA and MFA can be resolved. The infrared divergences in the loop
corrections to the four-fermion vertices are absorbed into the transition
form factors, which are not calculable in perturbation theory. The
external quarks then remain on-shell, and the problem of the scale
dependence is resolved without breaking the gauge invariance. Hence,
factorizable contribution in QCDF are treated in the same way as in MFA
[see eq.~(\ref{mfa})] but with different $C_{\rm eff}$.

Nonfactorizable contributions are calculated perturbatively in the heavy
quark limit. In the $B\to\pi\pi$ decays these contributions
are written as the convolutions of hard amplitudes with
meson distribution amplitudes $\phi$ in momentum fractions of valence
quarks,
\begin{eqnarray}
& &F^{B\pi}\otimes H^{(4)}\otimes \phi_{\pi 2}\;,
\\
& &\phi_B\otimes H^{(6)}\otimes \phi_{\pi 1}\otimes \phi_{\pi 2}\;,
\label{six}
\end{eqnarray}
where $H^{(4)}$ ($H^{(6)}$) represents a four-quark (six-quark)
amplitude. The former collects the infrared finite piece of the
corrections to the four-fermion vertices, and the latter corresponds to
the pair of nonfactorizable diagrams with a hard gluon emitted from the
spectator quark. However, in some cases the hard amplitudes contain
end-point singularities, which are not smeared by the meson distribution
amplitudes. To regulate these end-point singularities, cutoffs of the
momentum fractions need to be introduced, which are complex in 
general parametrization \cite{BBNS2}.

Annihilation diagrams have been neglected in FA. In QCDF these amplitudes
are calculated in a similar way to the nonfactorizable ones. The
end-point singularities still exist, and complex cutoffs must be
introduced. These complex cutoffs could bring in large strong phases
and large CP asymmetries in two-body nonleptonic $B$ meson decays,
since they are basically free parameters.

The BSS mechanism also contributes to the strong phases. In QCDF, because
of the introduction of meson distribution amplitudes, the external
momentum flowing through the charm loop can be defined rigorously.
Let the quark going into the pion emitted from the weak vertex carry
the momentum fraction $x_2$. The quark going into the pion involved in
the $B\to\pi$ transition carries the momentum fraction $x_3\sim 1$,
since the transtion form factor is assumed to be dominated by soft
dynamics. The invariant mass $q^2$ appearing in Eq.~(\ref{bss}) is then
expressed as $q^2=x_2 M_B^2$ unambigiously, $M_B$ being the $B$ meson
mass.

\section{Perturbative QCD}

The problems of FA and MFA are resolved in a different way in the
PQCD approach. The infrared divergences in the vertex corrections 
are treated in the presence of the spectator quark \cite{CLY}. Therefore,
the leading-twist $B$ meson (pion) wave function can be defined, which
absorbs the two-particle reducible infrared divergences on the $B$ meson
(pion) side. The two-particle irreducible infrared
divergences cancel between the pair of diagrams, for example, with the
gluon emitted from the $b$ quark attaching the light quark and the
spectator quark, which form the outgoing pion in the $B\to\pi$ transition.
In this treatment the external quarks also remain on-shell, and the
problem of the scale dependence is resolved without breaking gauge
invariance.

In the PQCD picture the hard amplitudes for various topologies of
diagrams, including factorizable, nonfactorizable and annihilation,
are all six-quark amplitudes \cite{CL,YL,WYL}. That is, the decay
amplitudes are written as the convolutions in Eq.~(\ref{six}). In PQCD
the end-point singularities do not exist because of the inclusion of
Sudakov effects \cite{LY1,L5}, and the arbitrary cutoffs in QCDF are not
necessary. Therefore, factorizable, nonfactorizable and annihilation
amplitudes can be estimated in a more consistent way in PQCD than in
QCDF. In QCDF factorizable contributions involve only four-quark
amplitudes. As explained later, this difference will lead to different
characteristic scales and different power counting rules in $1/m_b$,
$m_b$ being the $b$ quark mass, for two-body nonleptonic $B$ meson decays
in QCDF and in PQCD.

In PQCD strong phases mainly arise from the annihilation amplitudes,
which are almost imaginary \cite{KLS,LUY,KL}. The detailed reason is
referred to \cite{CKL}. The strong phases are large, since they appear
at the same order as the factorizable amplitudes. The BSS mechanism also
contributes to the strong phases. In terms of the notation in the
previous section, the invariant mass $q^2$ apearing in Eq.~(\ref{bss})
is expressed as $q^2=x_2 x_3M_B^2$ unambigiously. However, compared to
the annihilation contributions, the BSS mechanism is of next-to-leading
order, and less important.

\section{Sudakov Effects}

If calculating the $B\to\pi$ form factor $F^{B\pi}$ at large recoil using
the Brodsky-Lepage formalism \cite{BL,BTF}, a difficulty immediately
occurs. The lowest-order diagram for the hard amplitude is proportional to 
$1/(x_1 x_3^2)$, $x_1$ being the momentum fraction associated with the
spectator quark on the $B$ meson side. If the pion distribution amplitude
vanishes like $x_3$ as $x_3\to 0$ (in the leading-twist, {\it i.e.},
twist-2 case), $F^{B\pi}$ is logarithmically divergent. If the pion
distribution amplitude is a constant as $x_3\to 0$ (in the
next-to-leading-twist, {\it i.e.}, twist-3 case), $F^{B\pi}$ even becomes
linearly divergent. These end-point singularities have also appeared in
the evaluation of the nonfactorizable and annihilation amplitudes in QCDF
mentioned above.

In PQCD calculations small parton transverse momenta $k_T$ are included
\cite{LY1,LS}, which smear the end-point singularities from small momentum
fractions. Because of the inclusion of parton transverse momenta, double
logarithms $\ln^2(Pb)$ are generated from the overlap of collinear and soft
enhancements in radiative corrections to meson wave functions, where $P$
denotes the dominant light-cone component of a meson momentum, and $b$
is the variable conjugate to $k_T$. The resummation \cite{CS,BS} of these
double logarithms leads to a Sudakov form factor $\exp[-s(P,b)]$, which
suppresses the long-distance contributions from the large $b$ region with
$b\sim 1/\bar\Lambda$, $\bar\Lambda\equiv M_B-m_b$ representing a soft
scale. This suppression renders $k_T^2$ flowing into the hard amplitudes
of order
\begin{eqnarray}
k_T^2\sim O(\bar\Lambda M_B)\;.
\end{eqnarray}
The off-shellness of internal particles then remain of
$O(\bar\Lambda M_B)$ even in the end-point region, and the singularities
are removed. This mechanism is so-called Sudakov suppression.

Du {\it et al.} have studied the Sudakov effects in the evaluation
of nonfactorizable amplitudes \cite{Du}. If equating these amplitudes
with Sudakov suppression included to the parametrization in QCDF, it was
observed that the corresponding cutoffs are located in the reasonable
range proposed by Beneke {\it et al.} \cite{BBNS2}.
Sachrajda {\it et al.} have expressed an opposite opinion on the effect
of Sudakov suppression in \cite{GS}. However, their conclusion was drawn
based on a very sharp $B$ meson wave function, which is not favored by
experimental data.

It is easy to understand the increase of $k_T^2$ from $O(\bar\Lambda^2)$,
carried by the valence quarks which just come out of the initial meson
wave functions, to $O(\bar\Lambda M_B)$, carried by the quarks which are
involved in the hard weak decays. Consider the simple deeply inelastic
scattering of a hadron. The transverse momentum $k_T$ carried by a
parton, which just come out of the hadron distribution function, is
initially small. After infinite many gluon radiations, $k_T$ becomes of
$O(Q)$, when the parton is scattered by the highly virtual photon,
where $Q$ is the large momentum transfer from the photon. The evolution
of the hadron distribution function from the low scale to $Q$ is described
by the Dokshitzer-Gribov-Lipatov-Altarelli-Parisi (DGLAP) equation
\cite{AP}. The mechanism of the DGLAP evolution in DIS is similar to that
of the Sudakov evolution in exclusive $B$ meson decays. The difference
is only that the former is the consequence of the single-logarithm
resummation, while the latter is the consequence of the double-logarithm
resummation.

Another support for considering $k_T$ in the analyses of exclusive $B$
meson decays can be found in \cite{KPY}. It has been shown that the $k_T$
dependence appears as $\alpha_s\ln(1+k_T/k^+)$ in radiative corrections,
where $k^+$ ($k_T$) is the longitudinal (transverse) component of the
light spectator quark momentum. Obviously, the $k_T$ dependence does not
go away in the heavy quark limit, since both $k_T$ and $k^+$ are of
$O(\bar\Lambda)$.

\section{Power counting}

The power behaviors of various topologies of diagrams for two-body
nonleptonic $B$ meson decays with the Sudakov effects taken into account
has been discussed in details in \cite{CKL}. The relative importance is
summarized below:
\begin{eqnarray}
{\rm emission} : {\rm annihilation} : {\rm nonfactorizable} 
=1 : \frac{2m_0}{M_B} : \frac{\bar\Lambda}{M_B}\;,
\end{eqnarray}
with $m_0$ being the chiral symmetry breaking scale. The scale $m_0$
appears because the annihilation contributions are dominated by those
from the $(V-A)(V+A)$ penguin operators, which survive under helicity
suppression. In the heavy quark limit the annihilation and
nonfactorizable amplitudes are indeed power-suppressed compared to the
factorizable emission ones. Therefore, the PQCD formalism for two-body
charmless nonleptonic $B$ meson decays coincides with the factorization
approach as $M_B\to\infty$. However, for the physical value $M_B\sim 5$
GeV, the annihilation contributions are essential.

Note that all the above topologies are of the same order in $\alpha_s$
in PQCD. The nonfactorizable amplitudes are down by a power of $1/m_b$,
because of the cancellation between a pair of nonfactorizable diagrams,
though each of them is of the same power as the factorizable one. I
emphasize that it is more appropriate to include the nonfactorizable
contributions in a complete formalism. As stated in Sec. 2, the
factorizable internal-$W$ emisson contributions are strongly suppressed
by the vanishing Wilson coefficient $a_2$ in the $B\to J/\psi K^{(*)}$
decays \cite{YL}, so that nonfactorizable contributions become dominant.
In the $B\to D\pi$ decays, there is no soft cancellation between a pair
of nonfactorizable diagrams, and nonfactorizable contributions are
significant \cite{YL}.

In QCDF the factorizable and nonfactorizable amplitudes are of the same
power in $1/m_b$, but the latter is of next-to-leading order in
$\alpha_s$ compared to the former. Hence, QCDF approaches FA in the
heavy quark limit in the sense of $\alpha_s\to 0$. Briefly speaking,
QCDF and PQCD have different counting rules both in $\alpha_s$ and in
$1/m_b$. The former approaches FA logarithmically
($\alpha_s\propto 1/\ln m_b \to 0$), while the latter does linearly
($1/m_b\to 0$).

\section{Penguin Enhancement}

The leading factorizable contributions involve four-quark hard
amplitudes in QCDF, but six-quark hard amplitudes in PQCD. This
distinction also implies different characteristic scales in the two
approaches: the former is characterized by $m_b$, while the latter is
characterized by the virtuality of internal particles of order
$\sqrt{\bar\Lambda M_B}\sim 1.5$ GeV \cite{KLS,LUY,KL}. A six-quark hard
amplitude must contain a hard gluon exchanged between the spectator
quark and other quarks. The spectator quark in the $B$ meson, forming a
soft cloud around the heavy $b$ quark, carries momentum of order
$\bar\Lambda$. The spectator quark on the pion side carries momentum of
$O(M_B)$ in order to form the fast-moving pion with the light quark
produced in the $b$ quark decay. Based on this reasoning, the hard gluon
is off-shell by $O(\bar\Lambda M_B)$. As explored in \cite{L4}, this
scale, characterizing heavy-to-light decays, is important for
constructing a gauge-invariant $B$ meson wave function. The path-ordered
exponential in the definition of the $B$ meson wave function will appear,
only if the hard scale is of $O(\bar\Lambda M_B)$.

It has been known that to accommodate the $B\to K\pi$ and $\pi\pi$ data,
penguin contributions must be large enough. In FA, MFA and QCDF one
relies on chiral enhancement by increasng the mass $m_0$ to a large
value $m_0\sim 3$-4 GeV \cite{WS}. Because of the renormalization-group
evolution effect of the Wilson coefficients associated with the QCD
penguin operators, the lower hard scale leads to dynamical penguin
enhancement in PQCD. Whether dynamical enhancement or chiral enhancement
is responsible for the large $B\to K\pi$ branching ratios can be tested
by measuring the $B\to \phi K$ modes \cite{CKL,L6}. In these modes
penguin contributions dominate, such that their branching ratios are
insensitive to the variation of the unitarity angle $\phi_3$. Because
the $\phi$ meson is a vector meson, the mass $m_0$ is replaced by the
physical mass $M_\phi\sim 1$ GeV, and chiral enhancement does not exist.
If the branching ratios of the $B\to\phi K$ decays are around
$4\times 10^{-6}$ \cite{HMW,CY}, chiral enhancement may be essential for
the penguin-dominated decay modes. After including parametrized
annihilation contributions in QCDF, the $B\to\phi K$ branching ratios
reach around $7\times 10^{-6}$ at most \cite{CY}. If the branching ratios
are around $10\times 10^{-6}$ as predicted in PQCD \cite{CKL,M},
dynamical enhancement may be essential.
 
Recently, the charm penguin contributions \cite{S} have been proposed to
be mechanism alternative to chiral enhancement and dynamical enhancement.
It has been pointed out \cite{BG} that contributions from intrinsic
charms from the higher Fock states of the $B$ meson bound state may be
also essential for the explanation of the branching ratios and CP
asymmetries in the $B\to K\pi$ decays. These discussions indicate that
penguin contributions to two-body nonleptonic $B$ meson decays need more
thorough studies.

\section{Conclusion}

I have briefly reviewed the known approaches to two-body nonleptonic $B$
meson decays. Important aspects and phenomenological consequences of these
approaches have been discussed. It is an urgent mission to construct
a consistent and convincing approach to these decay modes. This requires
continuous confrontation between theoretical and experimental progresses.
After constructing such an approach, the Cabibbo-Kobayashi-Maskawa fit to
the data of two-body nonleptonic $B$ meson decays, as performed in
\cite{BBNS2}, will make more sense.

\section*{Acknowledgments}
I thank A. Ali, M. Beneke, S. Brodsky, G. Buchalla, M. Ciuchini, C.H. Chen,
H.Y. Cheng, M. Diehl, L. Dixon, D.S. Du, T. Feldmann, R. Fleischer,
S. Gardner, G. Hiller, T. Huang, Y.Y. Keum, G.P. Korchemsky, H. Lacker,
C.D. Lu, J.P. Ma, H. Quinn, L. Roos, C.T. Sachrajda, A.I. Sanda, L.
Silvestrini, G. Sterman,
Z.T. Wei, Y.L. Wu, K.C. Yang, M.Z. Yang, Z.X. Zhang and members in
the PQCD working group for useful
discussions. The work was supported in part by the National Science
Council of R.O.C. under the Grant No. NSC-89-2112-M-001-077, by the
National Center for Theoretical Science of R.O.C., and by Grant-in Aid
for Special Project Research (Physics of CP Violation) and by Grant-in
Aid for Scientific Exchange from Ministry of Education, Science and
Culture of Japan.

\end{document}